# Analysis and Design of a 32nm FinFET Dynamic Latch Comparator

Mir Muntasir Hossain
*Department of Electrical and Electronic Engineering*
*Ahsanullah University of Science and Technology*
Dhaka, Bangladesh
mirmuntasir.eee@gmail.com

Satyendra N. Biswas
*Department of Electrical and Electronic Engineering*
*Ahsanullah University of Science and Technology*
Dhaka, Bangladesh
sbiswas.eee@aust.edu

***Abstract*—Comparators have multifarious applications in various fields, especially used in analog to digital converters. Over the years, we have seen many different designs of single stage, dynamic latch type and double tail type comparators based on CMOS technology, and all of them had to make the tradeoff between power consumption and delay time. Meanwhile, to mitigate the short channel effects of conventional CMOS based design, FinFET has emerged as the most promising alternative by owning the tremendous gate control feature over the channel region. In this paper, we have analyzed the performance of some recent dynamic latch type comparators and proposed a new structure of dynamic latch comparator; moreover, 32nm FinFET technology has been considered as the common platform for all of the comparators circuit design. The proposed comparator has shown impressive performance in case of power consumption, time delay, power delay product and offset voltage while compared with the other recent comparators through simulations with LTspice.**

## I. Introduction

One of the most essential applications of comparators is in analog to digital converter design along with in other fields like line decoders of memory bits, level shifter in multiple voltage domains, data receivers, memory sense amplifiers, etc. [1]. Moreover, in bio-imaging applications, very high sensitivity, higher resolution and higher data rate based ADCs are currently very demanding [2].

Due to considerable amount of power consumption and restricted speed, static comparators are now being outdated and considered infeasible for current generations' sophisticated portable devices [3]. Some outstanding features like high input impedance, strong positive feedback, negligible static power consumption, rail to rail output swing, higher intrinsic gain have made dynamic latch comparators an attractive option [4]. Single stage comparators appeared promising, but they had to make compromise between energy requirements and offset voltage together with severe kickback noise issues due to interconnection between differential input stage and regeneration latch stage [5]. Also, there are limitations in total current flow through two output nodes because of a single tail transistor, which also enhances the dependence on offset voltage and speed for various ranges of input common mode voltage [6]. By providing the necessary isolation between the latch and preamplifier stage, double tail type architectures emerged as a new alternative of single stage comparators with near supply voltage function, higher input common mode voltage range, cascade amplifying stage, higher accuracy, etc. [7], [8]. However, precision in timing is required between the two separate clock signals because within a very limited time, the voltage difference of the first stage has to be detected. With the advancement in technology, including lower supply voltage, designing high speed comparators is more challenging [9]. In consequence, to overcome the lower supply voltage and offset mismatch issues, transistors with greater size are required thus causing more power consumption and chip area.

The most promising alternative scope of current CMOS based technology, to continue scaling, is FinFET which can be used to design low power and high speed dynamic latch comparators. Some great features of this multi-gate structure are higher carrier mobility, improved scalability, mitigation of conventional short channel effects, lower threshold and supply voltage, high frequency operation, reduced dopant fluctuations, lower level of leakage current at sub-threshold operation, superior channel control [10]. There has been very limited research so far carried out to design comparators using FinFET technology. Through this paper, we tried to overcome that gap and designed all of our comparators using 32 nm FinFET PTM models to provide a unique platform for comparison.

In this paper, for different architectures of dynamic comparators, several analysis have been carried out in order to demonstrate their performance contrasts. Moreover, a new dynamic latch comparator has been proposed which shows significant improvements in terms of power consumption, time delay, power delay product and offset voltage. The rest of the paper is organized as follows. Section II discusses the operation, merits and demerits of recent dynamic latch type comparators. In section III, our proposed new comparators' operation has been presented in details. Simulation results and performance comparison data are given in section IV. Lastly, the paper is concluded in section V.

## II. Dynamic Latch Comparators

In this section, we will discuss on the operation, pros and cons of existing dynamic comparators developed in recent times where each of them possess some unique characteristics.

The dynamic comparator presented in fig. 1 is designed by HeungJun Jeon et al. [11]. This comparator operates in two clock cycles. During the reset phase, when CLK=0, F4 and F5 P-FinFETs turn on while F1 remains off. The drain terminals of F4 and F5 are respectively connected to the inverter pairs of F16-F18 and F17-19. These inverters will produce logic 0. Thus F10, F11, F14 and F15 will turn on and charge Out+ and Out- terminals to Vdd and during this time F12 and F13 N-FinFETs will remain off. During evaluation phase when CLK=1, F4 and F5 will turn off and F1 will turn on and provide the path for discharging through F2 and F3. Depending on the voltage level of Vin and Vref this discharge will take place. If Vin is higher than Vref, the drain terminal charge of F4 will reach to ground earlier than drain terminal of F5. So, inverter F16-F18 will produce logic 1 and F12 will turn on and F14 and F10 will turn off. Consequently, Out- will be discharged to ground and turn on F9 which will produce logic 1 at the Out+ terminal and always keep F8 turned off. So, through latching, outputs will be remained fixed till again CLK=0 is applied for next cycle.

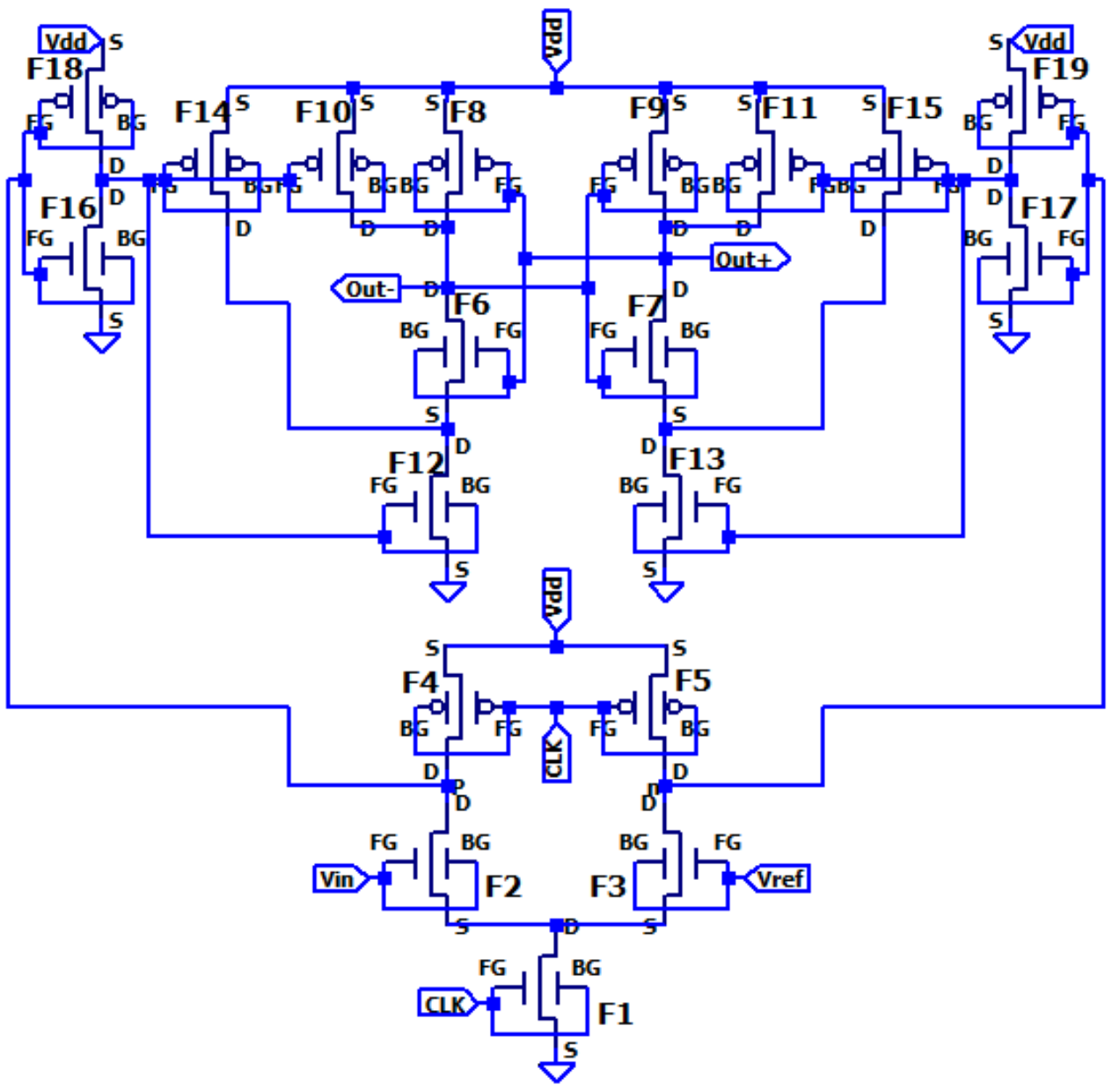

Fig. 1. Circuit diagram of the dynamic comparator presented in [11]

One of the key features is that here only a single clock has been used, so the clock precision performance has improved. Also, F14 and F15 have kept the drain terminals of F12 and F13 to Vdd, which has eventually produced faster response during evaluation phase. However, two inverter pairs consumed significant amount of power during the reset phase which is undesirable.

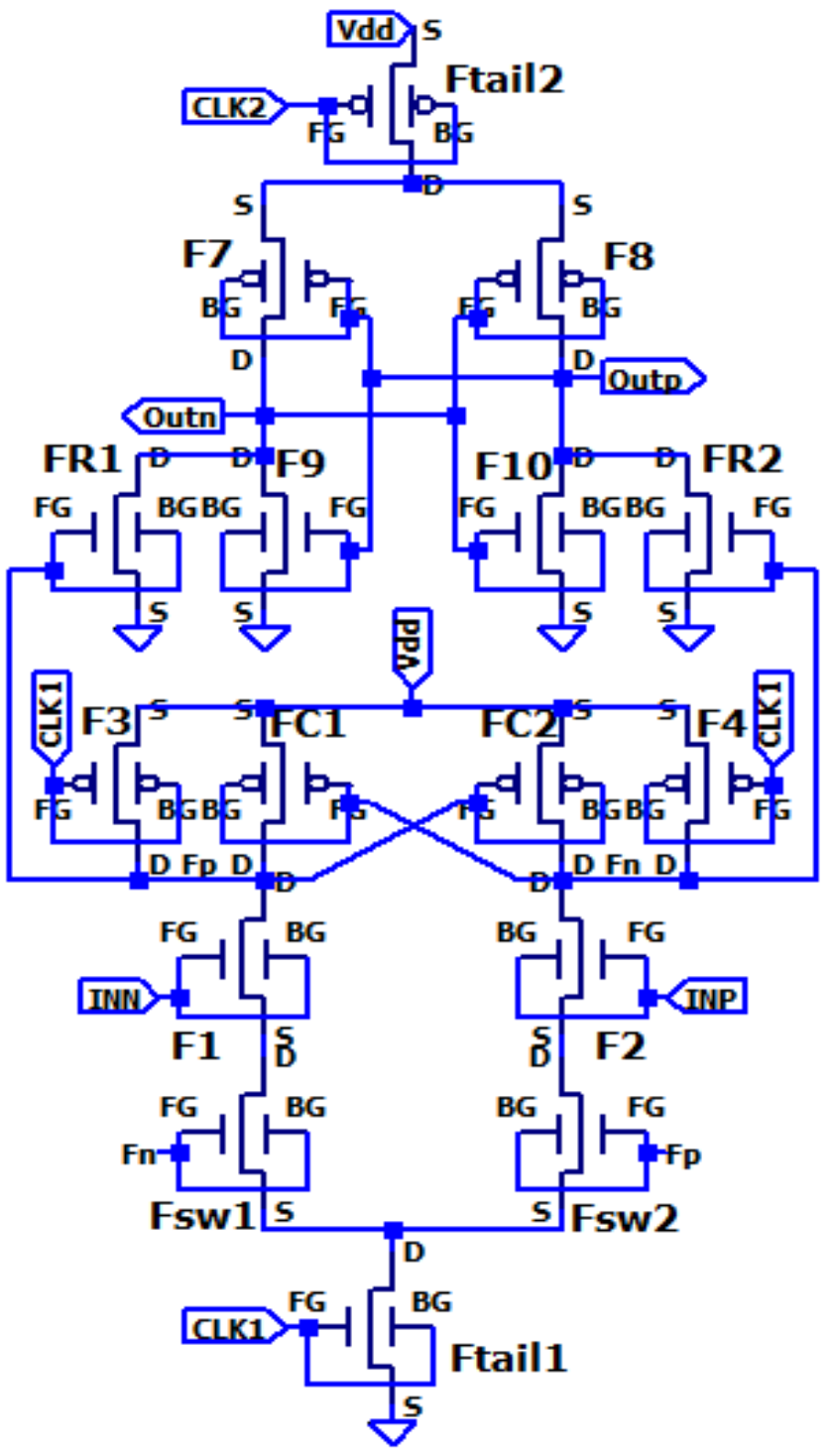

Fig. 2. Circuit diagram of the dynamic comparator presented in [12]

Another interesting double tail type dynamic comparator design is proposed by S. B. Mashhadi et al. [12] as shown in fig. 2. The operation of the circuit is as follows. When CLK1=0 and CLK2=1, F3 and F4 charge the Fn and Fp nodes to Vdd while the Ftail transistors remain off to prevent static power consumption. Moreover, as the gate of FR1 and FR2 are getting logic high they will discharge both of the output nodes to ground. During decision making phase, while CLK1=1 and CLK2=0, charge stored at Fn and Fp will start to discharge through Ftail1 depending on the voltage difference between INN and INP. Let’s assume that Fp is discharging faster than Fn. This will eventually turn on FC2, so Fn will have Vdd voltage and FC1 will be remained turn off and Fp will discharge to ground. Outp will also discharge to ground as the gate of FR2 is receiving logic high, and turn on F7 which will charge Outn to Vdd. Logic high at Outn will prevent F8 from turning on and charge Outp.

Additional inclusion in this circuit is Fsw1 and Fsw2. While Fp is discharging, Fsw2 is gradually turning off the discharging path of Fn and at the same time, Fsw1 is facilitating Fp to be completely discharged to the ground voltage. This feature is helping to reduce the dynamic power consumption during the switching phase.

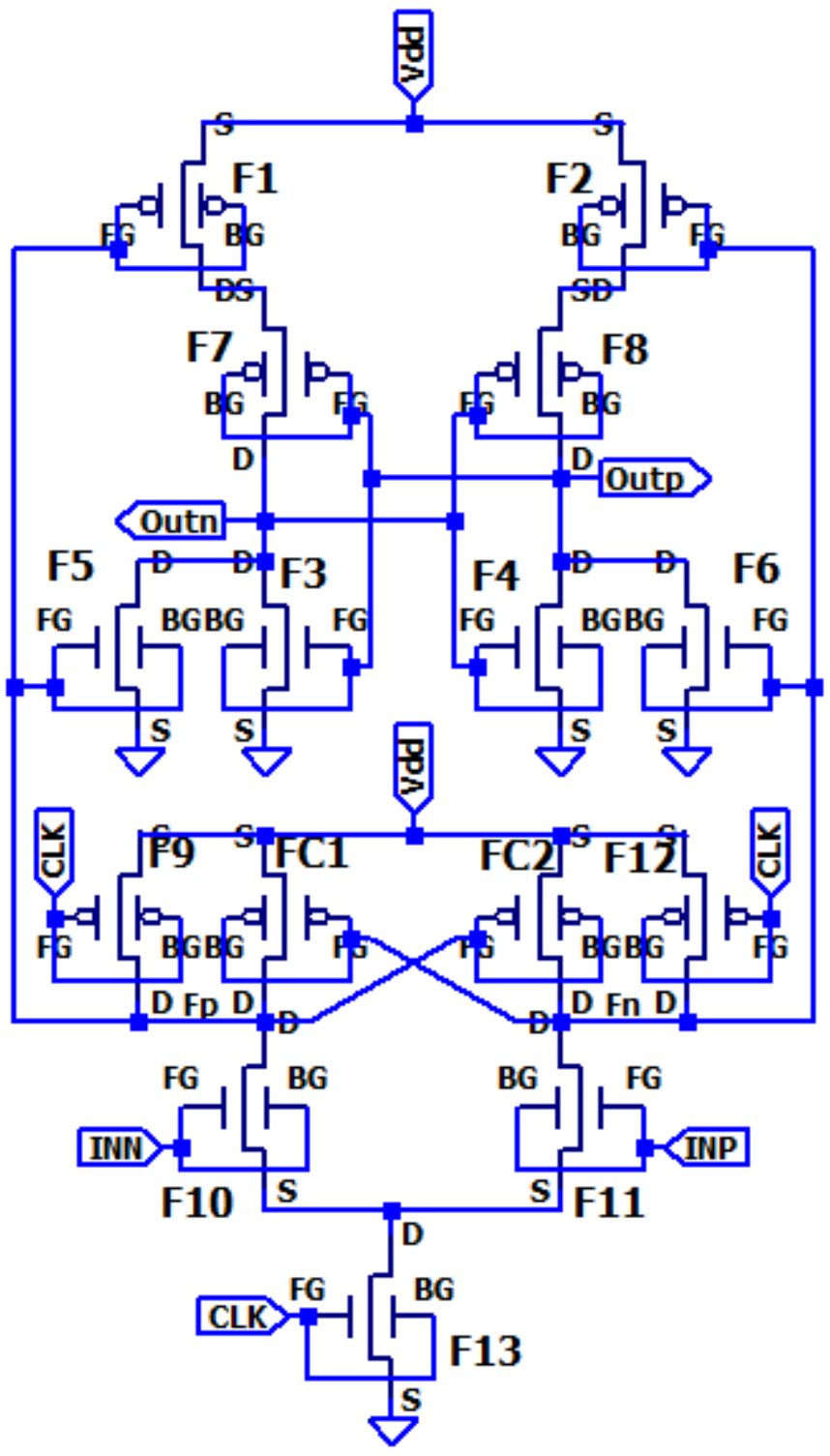

Fig. 3. Circuit diagram of the dynamic comparator presented in [13]

To avoid the coordination complexity of two different clock pulses, another type of dynamic comparator is designed by V. Deepika et al. [13], which is shown in fig. 3. The operation during the first stage remains same as the circuit of fig. 2, where Fn and Fp will be charged to Vdd when CLK=0 as F9 and F12 are on. During the comparison phase, depending on the conductance of F10 and F11, which is governed by the voltage level of INN and INP, Fn and Fp will start to discharge. If Fn discharges faster than Fp then eventually FC2 will be turned on and establish Fp to Vdd and Fp will continue to shut down FC1 thus Fn will subsequently become zero. Now, F1 and F6 will turn on and F2 and F5 will remain off. So, Outn becomes logic high and Outp logic low. Outp will keep F7 turned on and Outn will keep F8 turned off and establish the positive feedback latch.

The upper tail section of the conventional comparator is replaced by two p type transistors F1 and F2. So, the necessity of second clock has been diminished and sensitivity and gain of the second stage have also been improved. However, we found that these two transistors consumed substantial amount of power which eventually increased the average power requirement to a great extent.

## III. Proposed Dynamic Latch Comparator

In this section, we will discuss on our proposed design and its operation in details. Like other conventional comparators, this comparator also works into two separate phases. In the beginning, during the reset phase CLK1=0 and CLK2=1, so both Ftail1 and Ftail2 are off and through F5, F7 and F6, F8 both of the drain and source terminals of F3 and F4 are charged to Vdd. Outp and Outn are discharged to ground through F13 and F14 if they had any stored charge left from the previous cycle. Now, at the evaluation phase clock cycles are reversed and tail transistors are turned on. The conductance of path F1 and F2 depend on the voltage of Vin and Vref. Suppose, Vin is higher than Vref then node Fa will discharge faster than node Fb and , as F7 and F8 are now off, they will not have the scope to be recharged during this phase. Fa will turn off F4 and turn on F10, so node Fb will be at Vdd which will cause F9 off and F3 to turn on to facilitate Fa node to be completely discharged. Now, Outp will discharge to ground through F14 as node Fb will keep it on. Outp will also turn on F11 and turn off F15, so that Outn will be stayed at Vdd. Outn will cause F12 to turn off, so the charging path of Outp is now completely disconnected and turn on F16 to help Outp to be completely discharged.

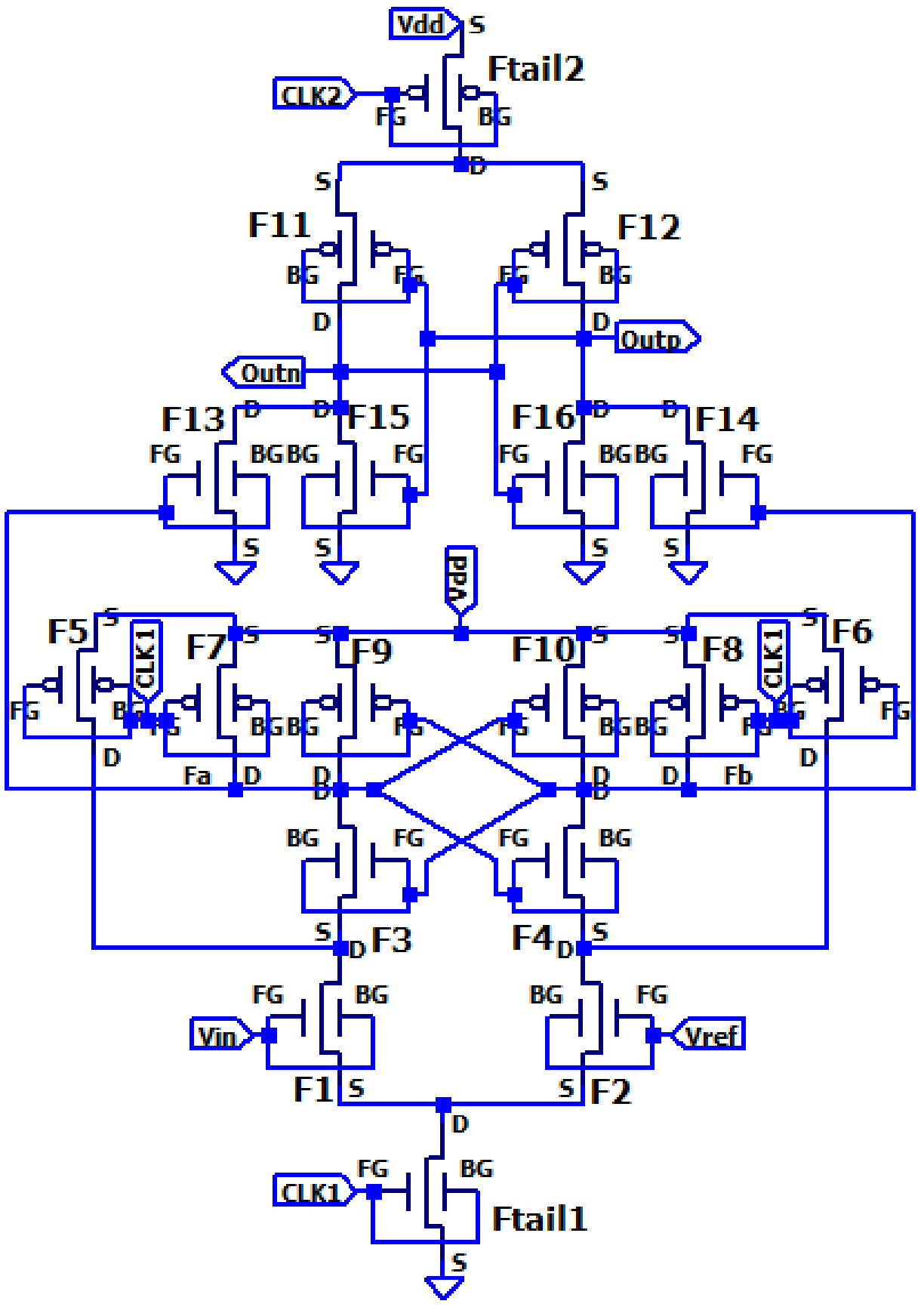

Fig. 4. Circuit diagram of the proposed dynamic latch comparator

In this design, F3 and F4 are helping to avoid static power consumption by disconnecting the discharging path after reaching decision. Also, F9 and F10 are causing one node to be recharged at Vdd level which is escalating the decision processing speed of Outp or Outn by providing higher gate voltage to F13 or F14. The top tail transistor is preventing any power consumption during reset phase. However, two separate clocks are needed and proper coordination between them is necessary to maintain the desired accuracy of comparison.

## IV. Simulation Results and Performance Analysis

All the dynamic latch comparators of fig. 1, fig. 2, fig. 3 and our proposed one, as shown in fig. 4, have been simulated with LTspice and plotted with MATLAB for better visibility. To show the proper comparisons among them, all the comparators are designed by using the 32nm FinFET PTM models which are developed by Arizona State University [14].

The clock cycle, input voltage, reference voltage, output voltage and average power consumption for all the comparators have been separately shown from fig. 5 to fig. 8. Form these figures, we can clearly observe the change in output voltage and time delay with respect to the change in clock cycle.

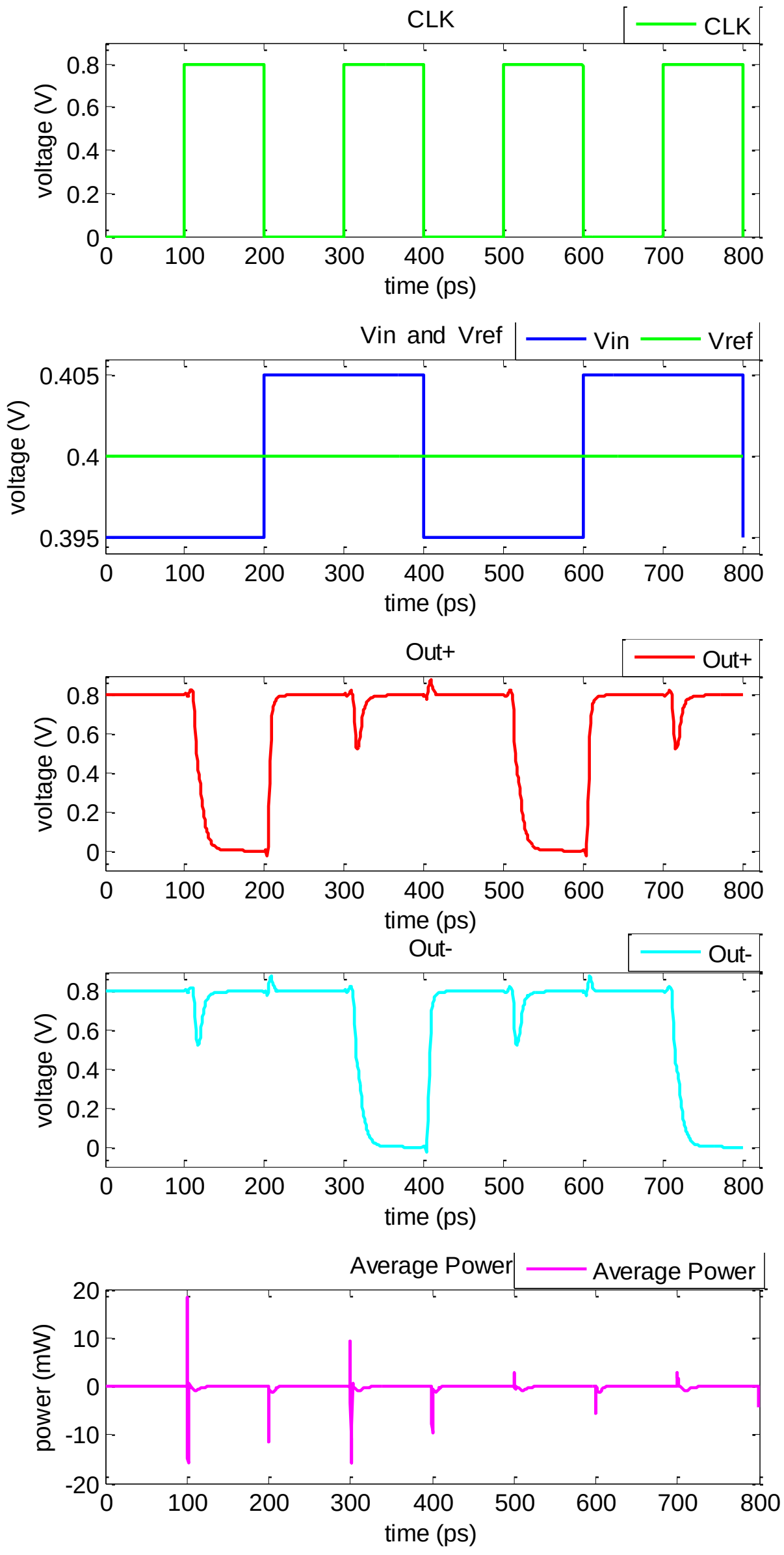

Fig. 5. Simulation results of the comparator of Ref [11]

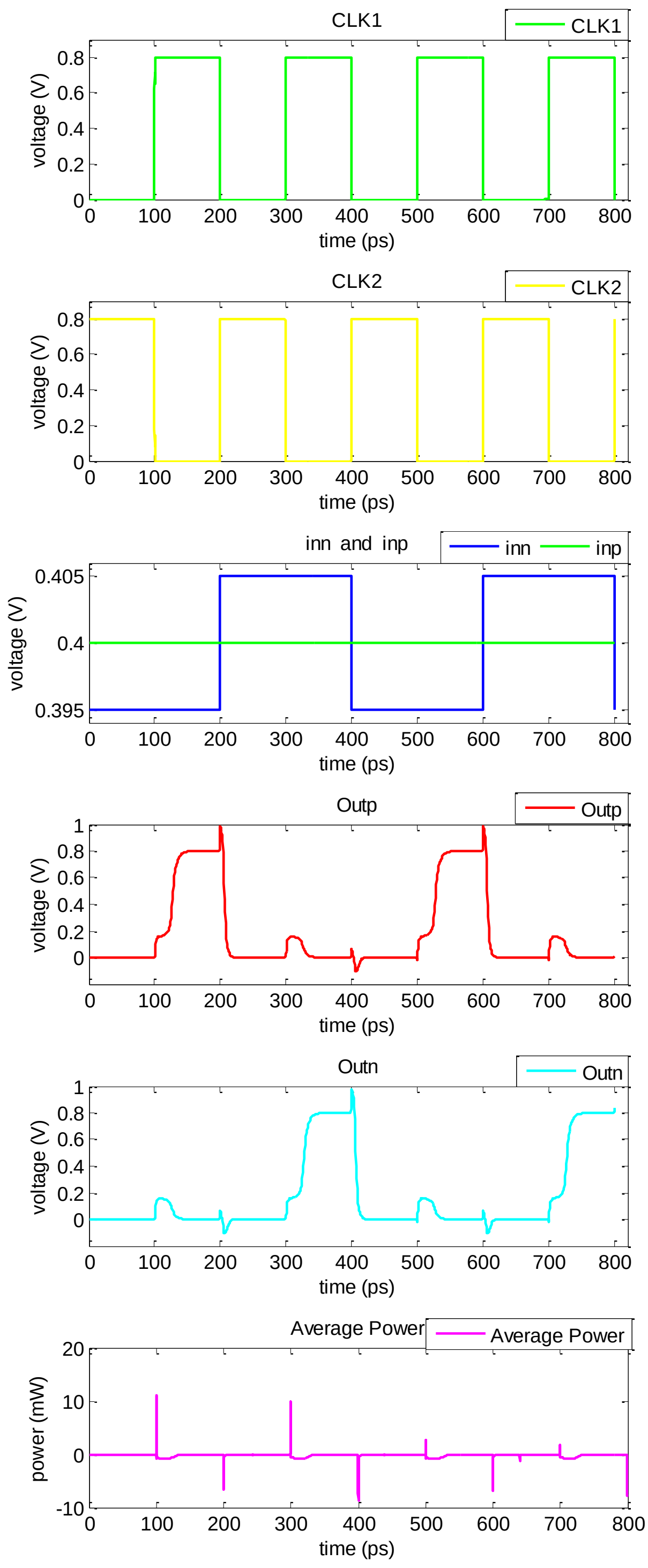

Fig. 6. Simulation results of the comparator of Ref [12]

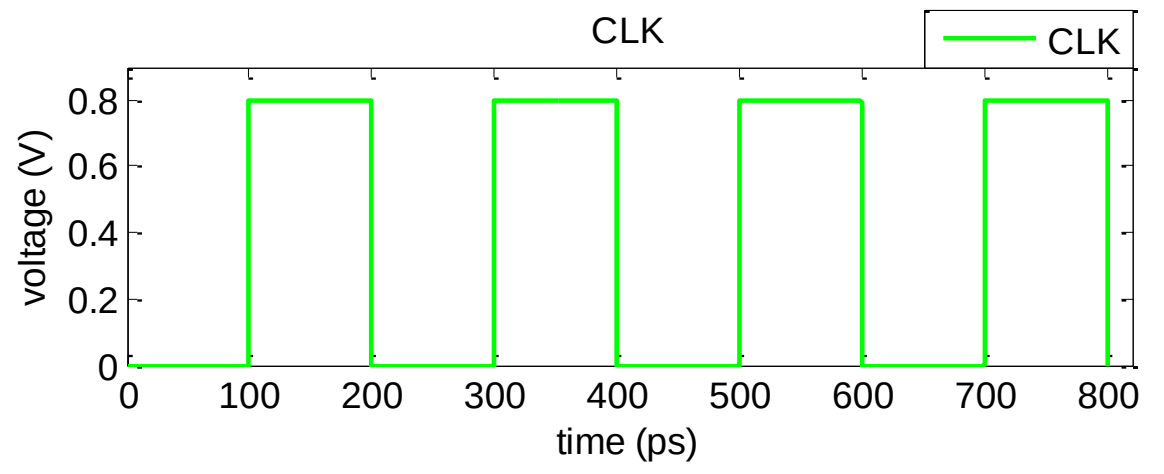

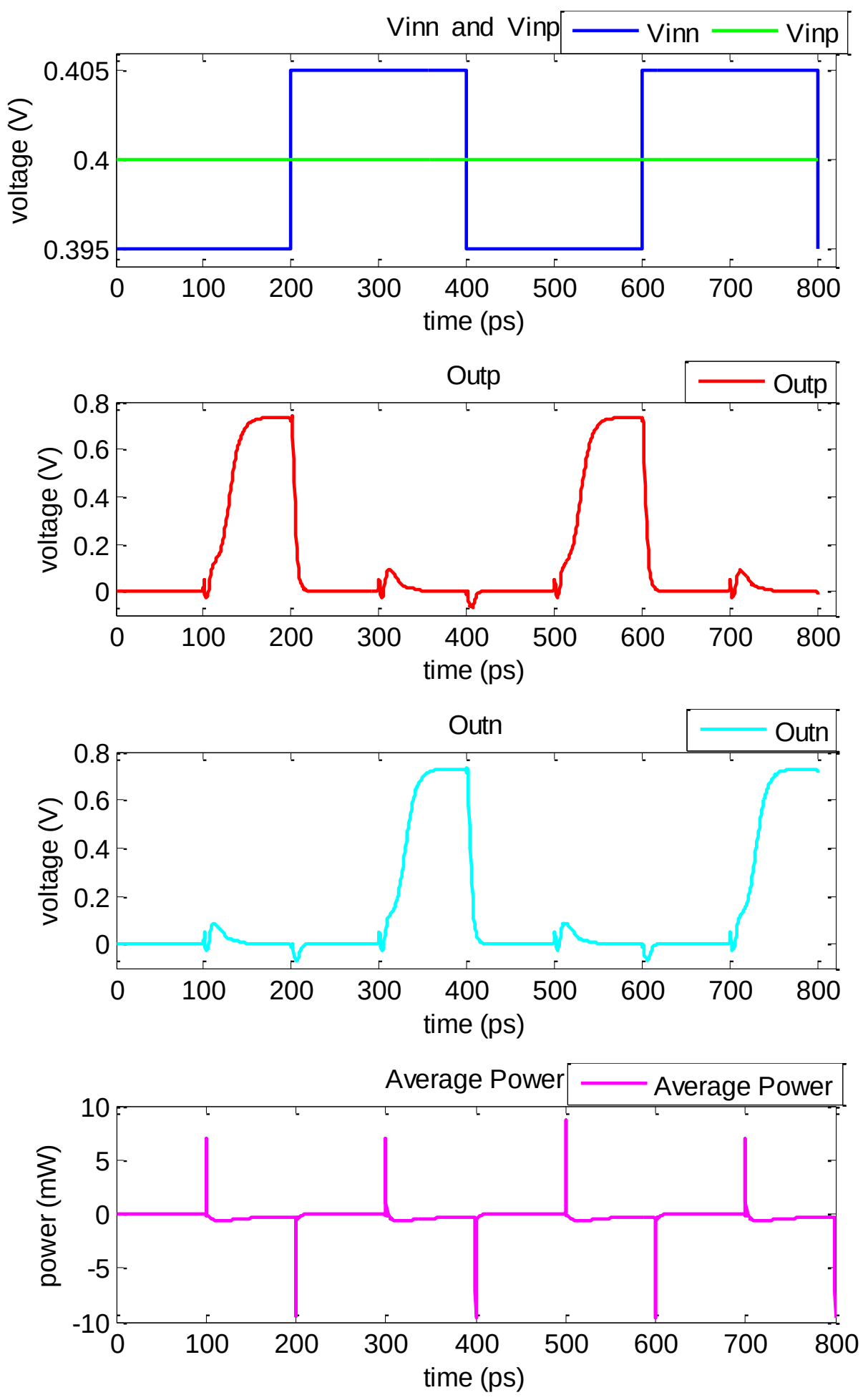

Fig. 7. Simulation results of the comparator of Ref [13]

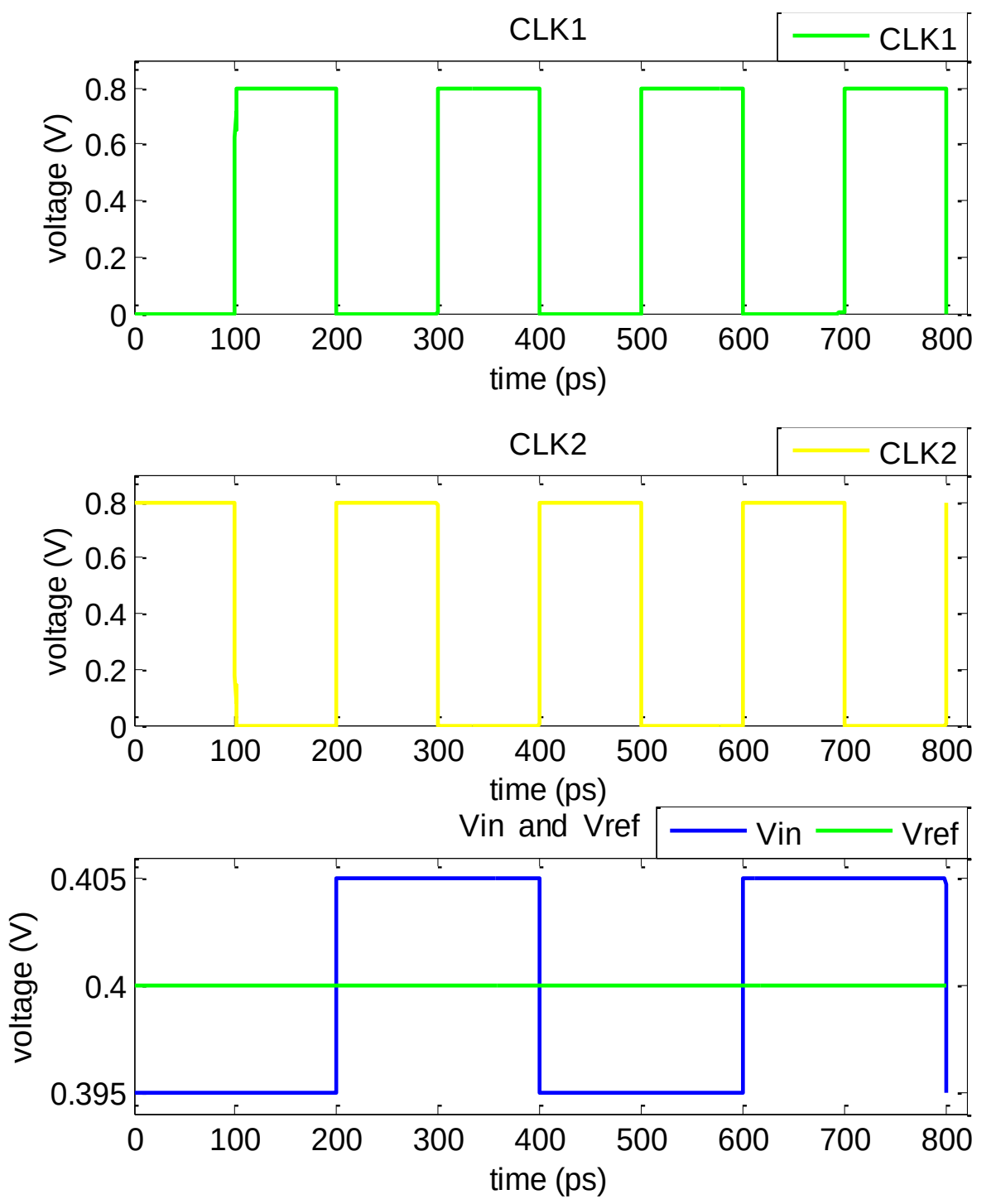

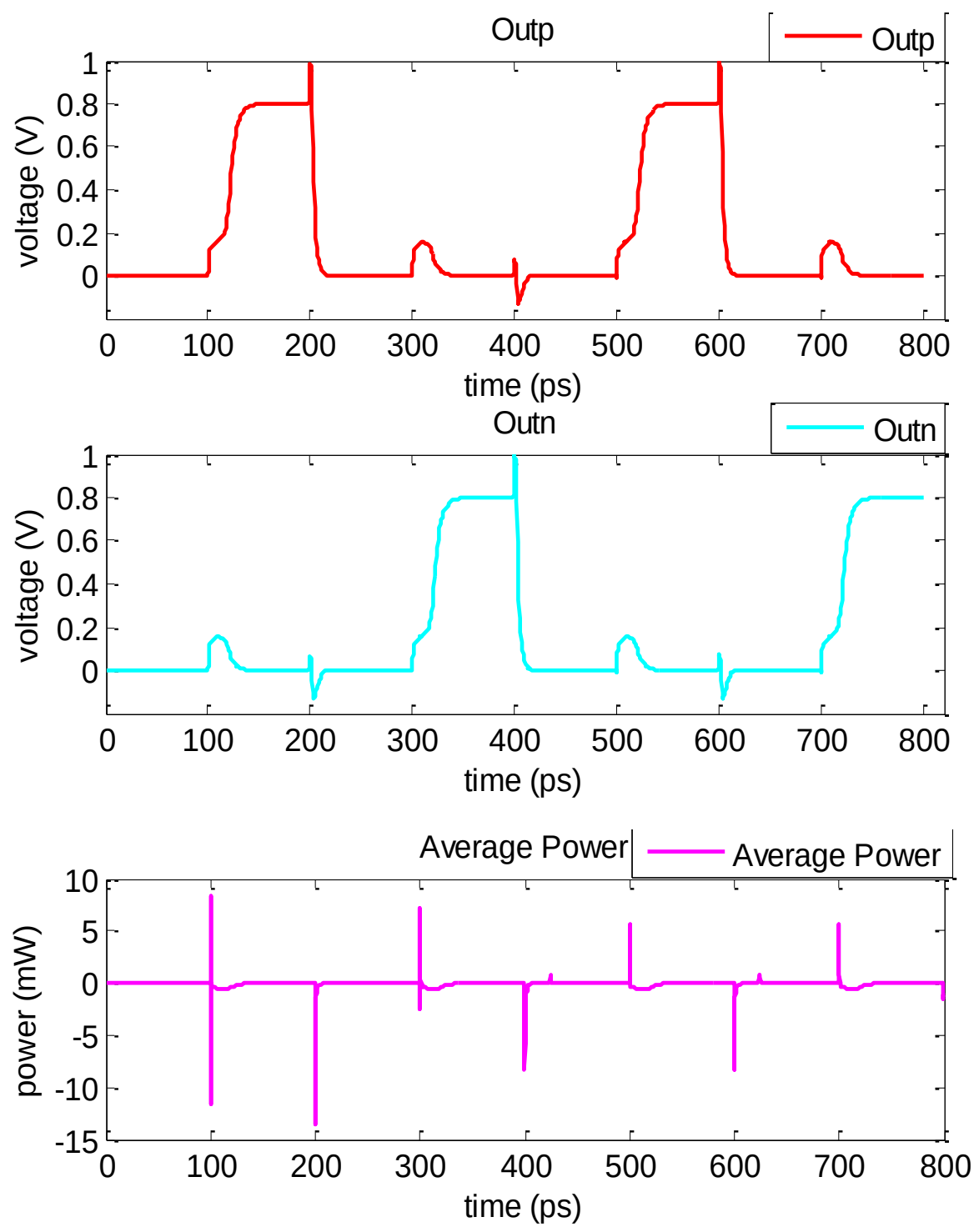

Fig. 8. Simulation results of the proposed comparator

While the voltage difference between the two inputs is varied from 5 mV to 100 mV, the delay time is significantly decreased for all the comparators. In fig. 9, the delay time for all the comparators with respect to their differential input voltages (ΔVin) is shown. The common mode voltage (Vcm) of 0.4 V is considered while ΔVin is varied. It has also been observed that the delay time of our proposed comparator is much smaller than other similar comparators.

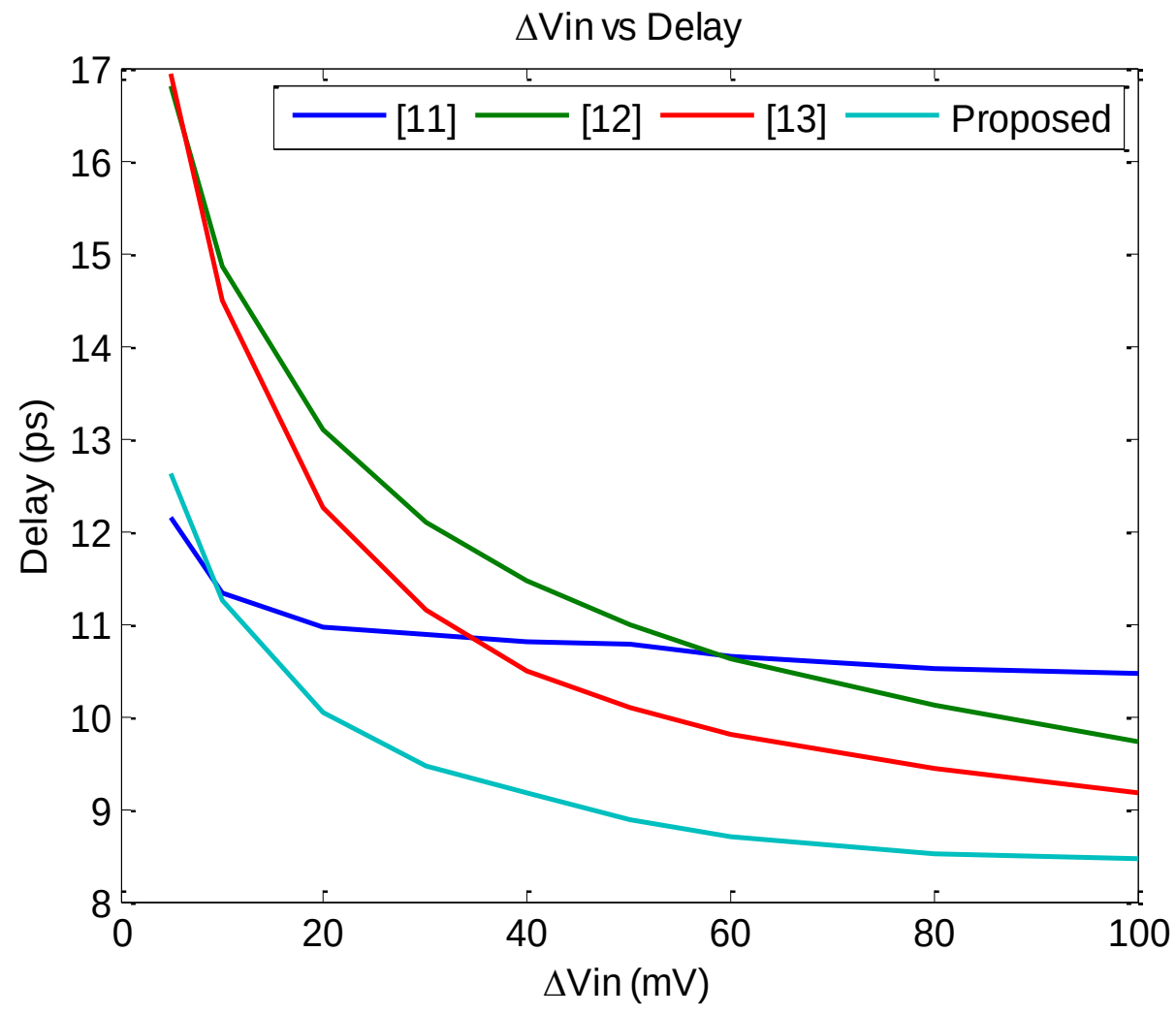

Fig. 9. ΔVin vs Delay analysis

In fig. 10, the delay time with respect to various input common mode voltage (Vcm) has been shown. As Vcm is increased from 0.3 V to 0.7 V, the delay time for all the comparators are decreased initially. But we have also noticed that at higher common mode voltage level, the delay time rises a little. Also, Vcm vs power consumption of various comparators is shown in fig. 11. The power consumption of our proposed comparator remains almost same for various common mode range. So, at different input levels, it will not cause sudden higher power demand which is clearly its one of the advantages. The differential input voltages (ΔVin) is chosen as 10 mV while Vcm is varied in this scenario.

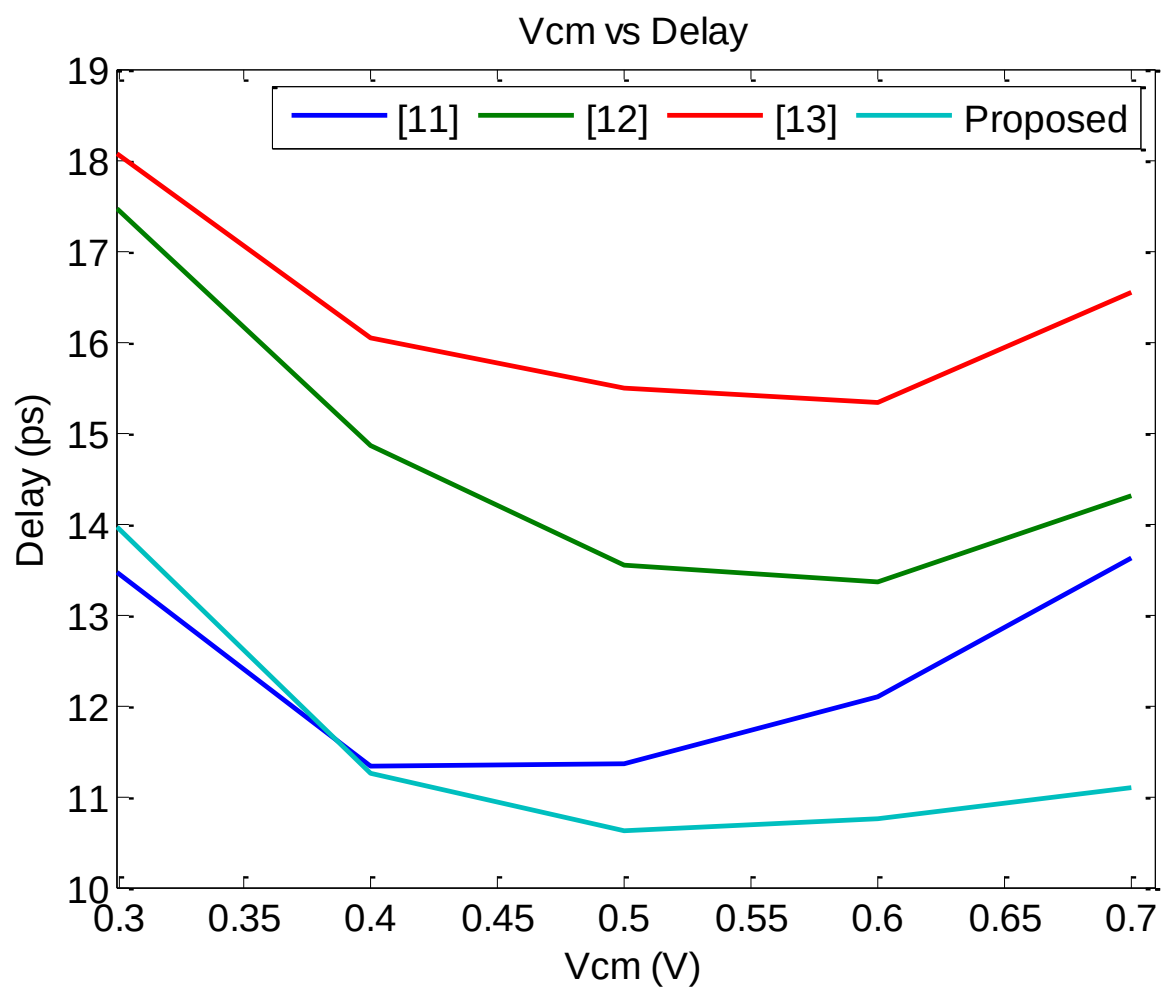

Fig. 10. Vcm vs Delay analysis

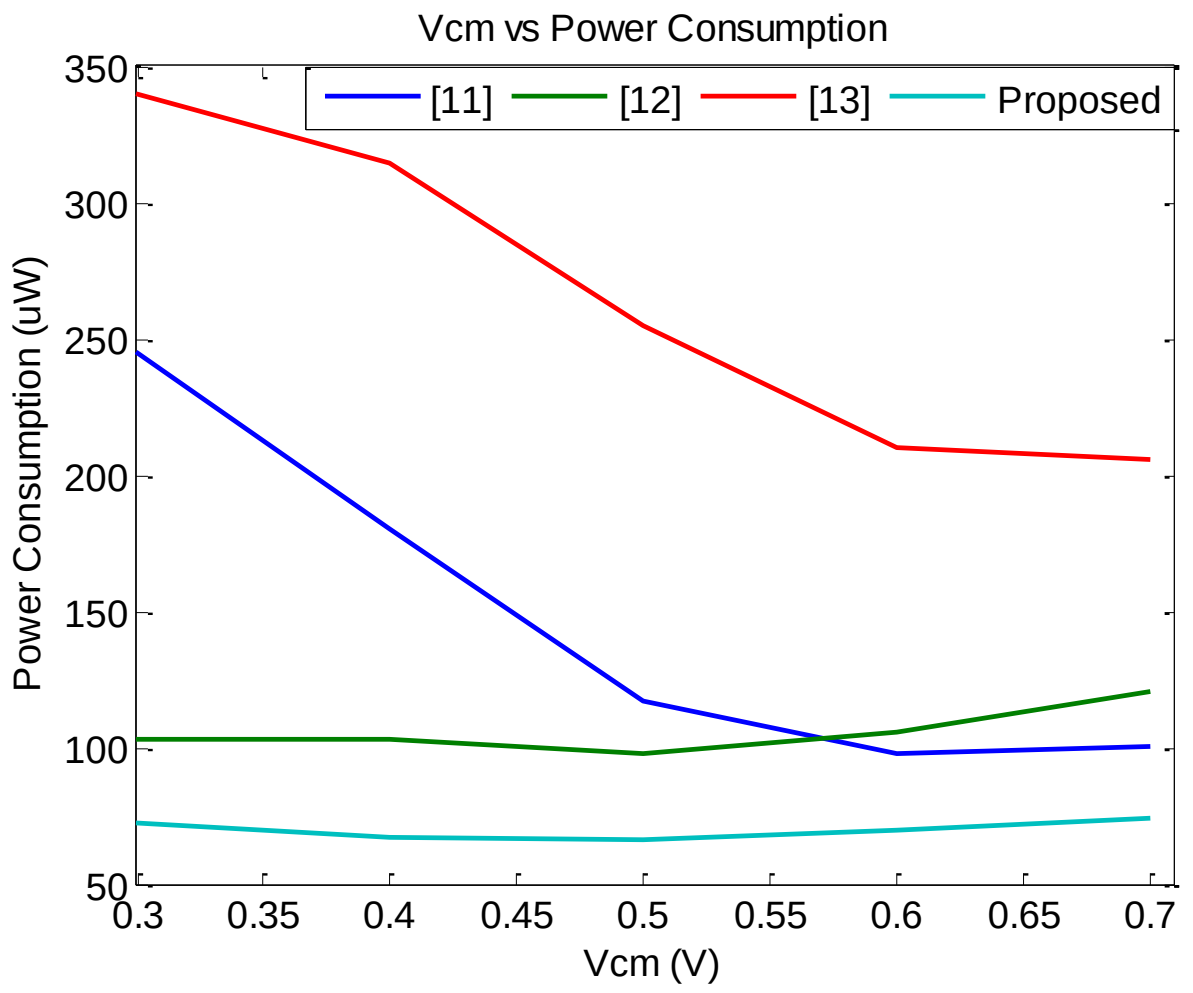

Fig. 11. Vcm vs Power Consumption analysis

The time delay at different supply voltages has been shown in fig. 12. As the FinFETs are designed to operate at lower voltage levels than CMOS, so at higher supply voltage level it takes more time to charge and discharge thus causing more time delay. Moreover, the power consumption for all of the comparators increased with the progression of the supply voltage which is quite obvious.

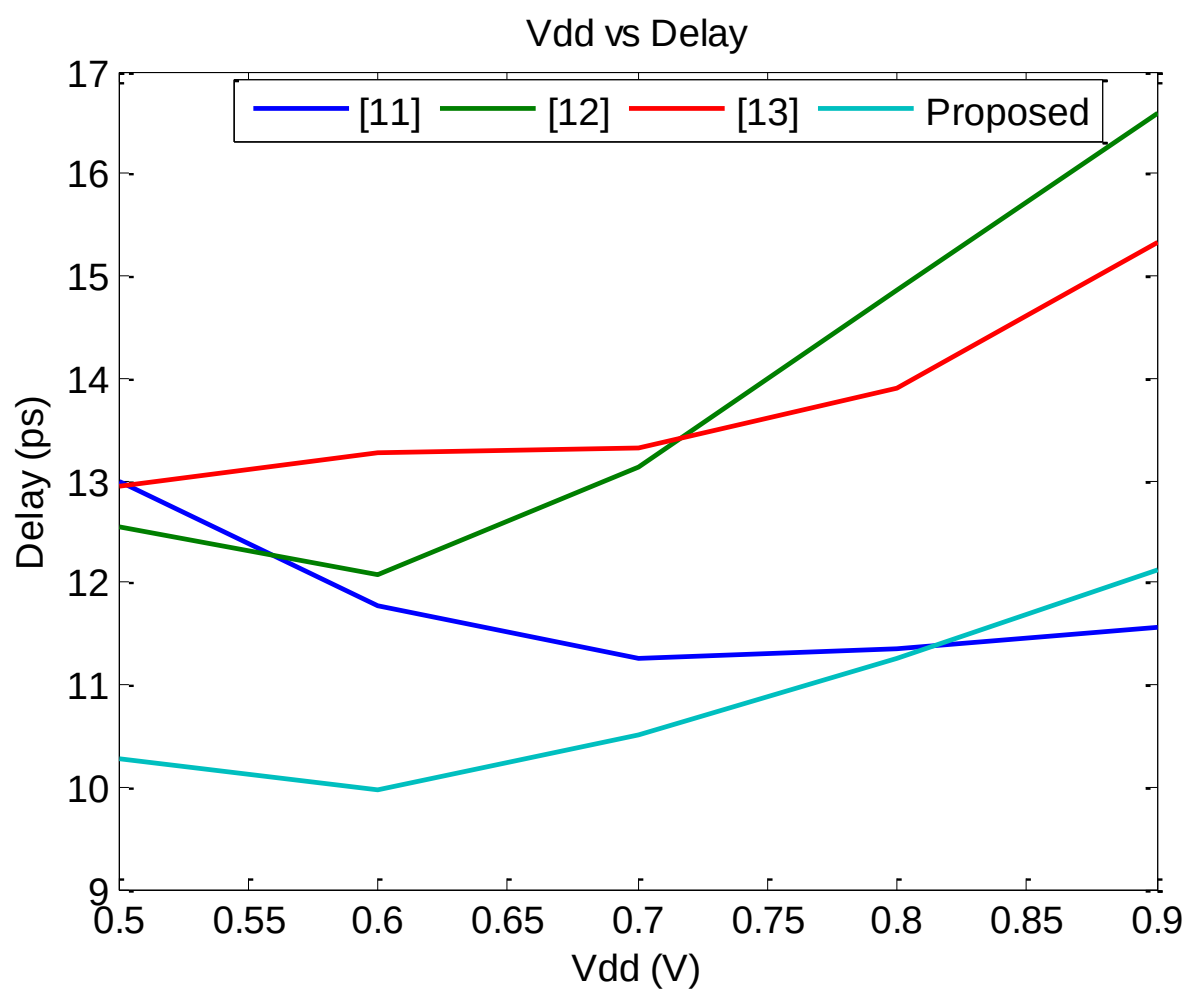

Fig. 12. Vdd vs Delay analysis

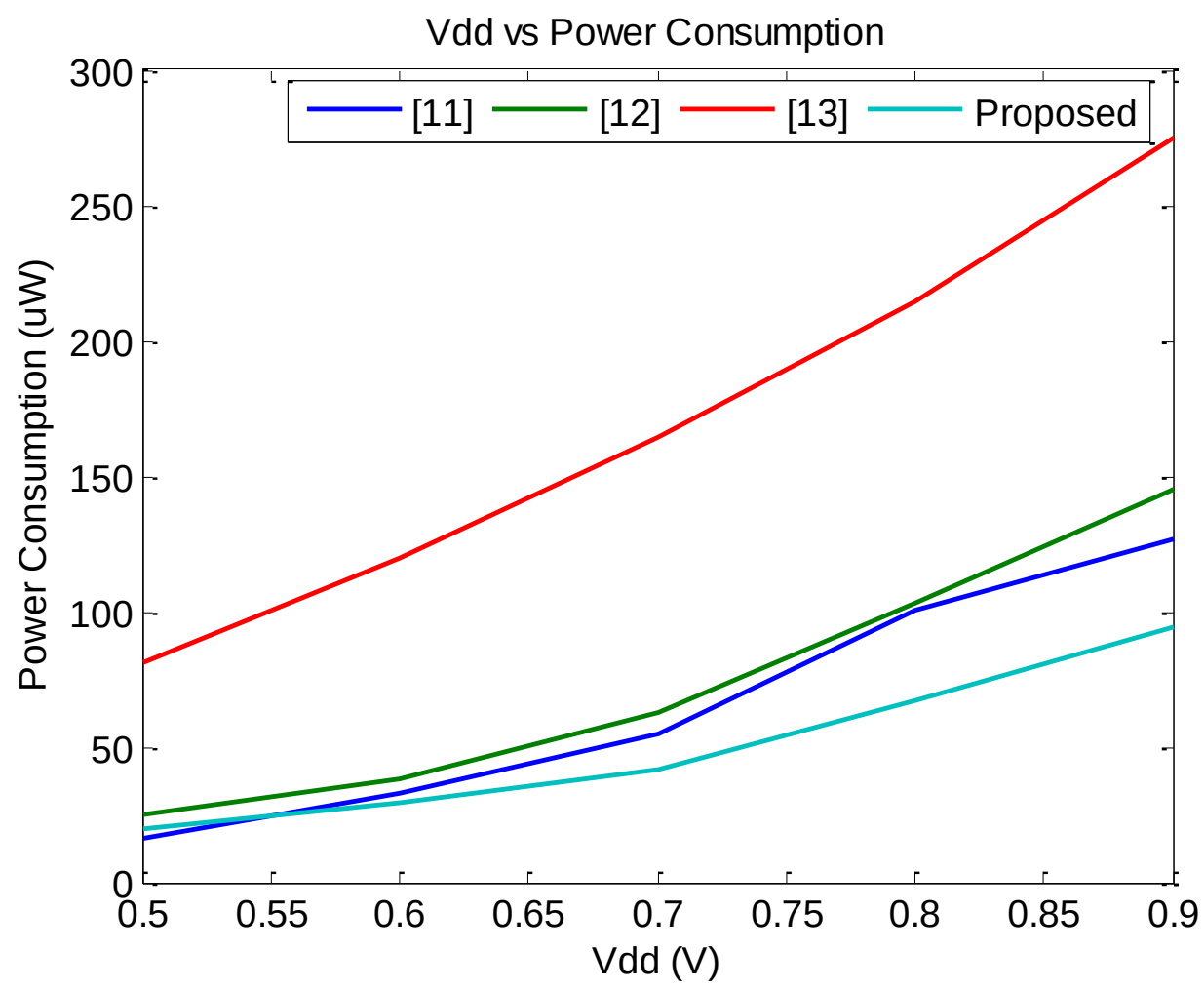

Fig. 13. Vdd vs Power Consumption analysis

The detailed performance comparisons of our proposed comparator with other similar works have been presented in Table I. The supply voltage, differential input voltage and operating frequency are held same at 0.8 V, 5 mV and 5 GHz, respectively for all of the comparators to show the proper contrasts. The average power consumption of our proposed comparator of 73.36 µW is the lowest among all of the structures. The propagation time delay of 12.63 ps of the suggested comparator is slightly higher than the lowest value of 12.15ps. Interestingly, from the power delay product comparisons, we can see that our proposed comparator consumes the least amount of energy for the fastest response which means in order to produce the same amount of delay, it will consume the minimum power. Lastly, the offset voltage level of our proposed comparator is found very reasonable of only 1.69 mV. Considering all the different measuring criteria's, we can say that the proposed comparator consumes less power together with faster response and a reasonable offset voltage.

TABLE I. PERFORMANCE COMPARISONS WITH OTHER DYNAMIC LATCH COMPARATORS

| Parameters | [11] | [12] | [13] | Proposed |
|---|---|---|---|---|
| Technology | 32nm FinFET | 32nm FinFET | 32nm FinFET | 32nm FinFET |
| Supply Voltage (V) | 0.8 | 0.8 | 0.8 | 0.8 |
| ΔVin (mV) | 5 | 5 | 5 | 5 |
| Clock (GHz) | 5 | 5 | 5 | 5 |
| Power Consumption (µW) | 150.11 | 114.47 | 222.18 | 73.36 |
| Propagation Delay (ps) | 12.15 | 16.81 | 16.93 | 12.63 |
| Offset Voltage (mV) | 1.55 | 1.82 | 3.34 | 1.69 |
| Power Delay Product (fJ) | 1.82 | 1.92 | 3.76 | 0.926 |

## V. CONCLUSION

In this paper, we have presented the design and analysis of a new dynamic latch type comparator. To design the proposed circuit, 32 nm FinFET PTM models are used and its competency has been verified through simulation results which are performed with LTspice. The average power consumption, time delay, offset voltage and power delay product results are found very promising and shown the proposed comparator's supremacy while compared with other recent works, designed using the same technology.